\documentclass[authoryear]{elsarticle}

\usepackage{amsmath}
\usepackage{amssymb}
\usepackage{amsthm}
\usepackage{natbib}
\usepackage{wasysym}
\usepackage{graphicx}
\usepackage{subcaption}
\usepackage{float}
\usepackage{epsfig}
\usepackage{booktabs}
\usepackage{multirow}
\usepackage{lineno}
\usepackage{slashbox}
\usepackage{etoolbox}
\usepackage[margin=4cm]{geometry}

\begin{document}

\title{XUV-driven mass loss from extrasolar giant planets orbiting active stars}

\author[icl]{J.M.~Chadney\corref{cor}}
\ead{joshua.chadney10@imperial.ac.uk}

\author[icl]{M.~Galand}

\author[icl]{Y.C.~Unruh}

\author[lpl]{T.T.~Koskinen}

\author[csic]{J.~Sanz-Forcada}

\cortext[cor]{Corresponding author}

\address[icl]{Department of Physics, Imperial College London, Prince Consort Road, London SW7 2BW, UK}

\address[lpl]{Lunar and Planetary Laboratory, University of Arizona, 1629 E. University Blvd., Tucson, AZ 85721, USA}

\address[csic]{Centro de Astrobiolog\`{i}a (CSIC-INTA), ESAC Campus, P.O. Box 78, E-28691 Villanueva de la Ca\~{n}ada, Madrid, Spain}

\begin{abstract}
Upper atmospheres of Hot Jupiters are subject to extreme radiation conditions that can result in rapid atmospheric escape. The composition and structure of the upper atmospheres of these planets are affected by the high-energy spectrum of the host star. This emission depends on stellar type and age, which are thus important factors in understanding the behaviour of exoplanetary atmospheres. In this study, we focus on Extrasolar Giant Planets (EPGs) orbiting K and M dwarf stars. XUV spectra for three different stars –- $\epsilon$ Eridani, AD Leonis and AU Microscopii –- are constructed using a coronal model. Neutral density and temperature profiles in the upper atmosphere of hypothetical EGPs orbiting these stars are then obtained from a fluid model, incorporating atmospheric chemistry and taking atmospheric escape into account. We find that a simple scaling based solely on the host star's X-ray emission gives large errors in mass loss rates from planetary atmospheres and so we have derived a new method to scale the EUV regions of the solar spectrum based upon stellar X-ray emission. This new method produces an outcome in terms of the planet's neutral upper atmosphere very similar to that obtained using a detailed coronal model of the host star. Our results indicate that in planets subjected to radiation from active stars, the transition from Jeans escape to a regime of hydrodynamic escape at the top of the atmosphere occurs at larger orbital distances than for planets around low activity stars (such as the Sun).
\end{abstract}

\maketitle

\section{Introduction}
In order to properly understand what exoplanets are made of and thus infer their formation and evolution, one needs information on their atmospheres. In recent years, transit spectroscopy has been used to derive absorption and emission spectra of exoplanetary atmospheres. The first detection of a planetary atmosphere was that of the Hot Jupiter HD209248b, at which \citet{Charbonneau2002} observed a dimming in the Na D lines during transit. The planet was subsequently found to possess an extended hydrogen cloud \citep{Vidal-Madjar2003} that is escaping hydrodynamically \citep{Koskinen2010,Koskinen2013a,Koskinen2013b}. Other close-orbiting extrasolar gas giants, such as HD189733b \citep{LecavelierdesEtangs2010} and WASP-12b \citep{Fossati2013}, also possess extended atmospheres that are most likely escaping hydrodynamically, due to the extreme radiation environments in which they are located. The upper atmospheres of extrasolar planets have been the subject of significant modelling effort, to help interpret the scarce observations (e.g., \citet{Yelle2004,GarciaMunoz2007,Koskinen2007,Koskinen2007a,Penz2008,Tian2008,Tian2008a,Koskinen2013a,Koskinen2013b,Owen2013}). These studies predict that planets orbiting at very small distances from their host stars -- HD209458b, for example, has an orbital distance of 0.047~AU -- should have escaping atmospheres. In particular, \citet{Koskinen2007,Koskinen2007a} found that hydrodynamic escape sets in once the stellar XUV\footnote{Note that in this paper, the following photon wavebands are used: X-ray (0.517 -- 12.4~nm), Extreme Ultraviolet EUV (12.4 -- 91.2~nm), Far Ultraviolet FUV ($\sim$90 -- 200~nm) (unless otherwise specified). XUV is used to refer to the combined X-ray and EUV wavelength ranges.} flux incident on the planet is strong enough to dissociate the main molecules responsible for cooling the upper atmosphere. For instance, the dissociation of H$_2$ means the infrared (IR) coolant H$_3^+$ cannot be formed. They predicted that this would be the case for Jupiter-like gas-giant planets orbiting within about 0.2~AU from a star of similar age and spectral type to the Sun.

These previous studies have not examined the influence of the high energy spectral shape of low-mass stars other than the Sun on the atmospheres of EGPs. In particular, current observing efforts are being focused more and more on K and M dwarfs \citep{LecavelierdesEtangs2012,Kulow2014}. M dwarfs are particularly interesting since they are the most common star type in our galaxy and present advantageous star-to-planet size ratios for transit spectroscopy. Additionally, K and M stars have lower effective temperatures than Sun-like G stars, resulting in Habitable Zones (HZs) located at smaller orbital distances \citep{Kasting1993}, which also increases chances of detecting habitable worlds.

The high-energy radiation environment of low-mass stars has been studied in recent years. \citet{France2013} describe FUV and NUV (170 -- 400~nm) radiation of a sample of 6 exoplanet-hosting M dwarfs. All of these stars are active in UV wavelengths and have very different spectral shapes to that of solar-like stars. Indeed, the ratio FUV/NUV is found to be around $10^3$ times higher in M dwarfs than in the Sun. This is due to lower NUV fluxes in the cooler M stars, as well as higher Lyman $\alpha$ line intensities.

\citet{Shkolnik2014} also study the FUV and NUV environment of early M stars, focussing in particular on the time-evolution of stellar irrandiances. The authors find that UV radiation remains at a saturated level in very young stars (up to a few 100~Myr) before declining as the stars age, with shorter wavelengths undergoing faster reductions in flux levels. This is similar behaviour to what has previously been found for solar-like stars \citep{Ribas2005} and for X-ray wavelengths in low-mass stars \citep{Sanz-Forcada2011}. \citet{Shkolnik2014} also note that X-ray and UV fluxes correlate over a broad range of stellar activity levels.

\citet{Linsky2014} derive scaling laws for the unobservable portion of the EUV waveband, based on chromospheric lines, such as Lyman $\alpha$. These complement the coronal model of \citet{Sanz-Forcada2011}, which is based on Emission Measure Distributions (EMDs) of the stellar atmospheres. The EMDs are determined using measured intensities of stellar emission lines in the X-ray, EUV and FUV; these are emissions emanating from the corona, transition region and chromosphere.

In this work, we focus on understanding how different high energy stellar emissions from low-mass stars of various ages and spectral types affect the properties of upper planetary atmospheres, including atmospheric escape. We use the coronal models of \citet{Sanz-Forcada2011} to obtain XUV spectra for three low-mass stars: $\epsilon$ Eridani, AD Leonis and AU Microscopii (see Section~\ref{sec:xexo}).  This is the first time that realistic stellar spectra have been used for thermospheric studies. We derive a more effective scaling method for the solar spectrum to be used when studying planets orbiting active stars (see Section~\ref{sec:scalings}).

We provide predictions of mass loss rates from EGP atmospheres (see Section~\ref{sec:escape}) and show how these are influenced by the spectral shape of the stellar XUV radiation and its intensity. We note that the current transit observations cannot be used to directly infer the mass loss rate.  This is because poor signal-to-noise of the observations, combined with stellar variability, means that all existing mass loss estimates are model-dependent \citep{Ben-Jaffel2007,Koskinen2010,Ben-Jaffel2013}.

\section{Stellar spectra} \label{sec:stellar_spec}
\subsection{Observational difficulties in the XUV} \label{sec:obs_diff}
Absorption of stellar radiation in planetary thermospheres (considered here to be the region above $p=1\mu\text{bar}$) occurs mainly to photons in the X-ray and EUV bands. Whilst it is possible to observe stellar emissions in the X-ray part of the spectrum, observing them in the EUV is either difficult or impossible. Indeed, at wavelengths greater than about 40~nm (and below 91.2~nm -- the H ionisation threshold), stellar radiation is almost completely absorbed by the inter-stellar medium (ISM), even for nearby stars. Furthermore, there are no current or planned missions to measure the observable portion of the EUV spectrum ($\lambda\lesssim 40$~nm) and we must rely on a limited number of old observations from the EUVE spacecraft. Therefore, to properly characterise the heating of exoplanetary atmospheres, we use a stellar coronal model \citep{Sanz-Forcada2011} to produce XUV spectra for three young, active, low-mass stars. The coronal model -- described in Section \ref{sec:xexo} -- is calibrated using observed XUV and FUV emission line intensities; so the stars chosen are close, bright objects, with good signal-to-noise observations from instruments like Chandra, XMM-Newton, ROSAT, EUVE, FUSE and IUE (see Table~\ref{tab:instr}). These stars are the K-dwarf $\epsilon$ Eridani and the M-dwarfs AD Leonis and AU Microscopii, some of the properties of which are provided in Table \ref{tab:stellar_prop}. 

\begin{table}
\centering
\caption{Wavelength range, and bin width ($\Delta\lambda$) or resolution ($R$) of instruments used to observe solar and stellar fluxes in the soft X-ray, EUV and FUV.}
\begin{tabular}{lcc}
\toprule
 & \textbf{$\lambda$ range} & $\Delta\lambda$ [nm] or \\
 & [nm] & $R = \lambda/\Delta\lambda$ \\
 \midrule
 \multicolumn{3}{c}{\textit{\textbf{Solar observatory}}} \\
 TIMED/SEE$^1$ & 0.5 -- 190 & $\Delta\lambda=$ 0.4 -- 7 \\
 \midrule
 \multicolumn{3}{c}{\textit{\textbf{Stellar observatories}}} \\
 Chandra/LETG$^2$ & 0.6 -- 15 & $\Delta\lambda=0.005$ \\
 Chandra/HETG$^2$ & 0.12 -- 3.1 & $\Delta\lambda=$ 0.0012 -- 0.0023 \\
 XMM-Newton/RGS$^3$ & 0.5 -- 3.5 & $R=$ 100 -- 500 \\
 ROSAT/PSPC$^4$ & 0.5 -- 12.4 & X \\
 EUVE$^5$ & 7 -- 76 & $R=$ 250 -- 500 \\
 FUSE$^6$ & 92 -- 118 & $R=20000$ \\
 IUE/SWP$^7$ & 120 -- 200 & $R=300$ \\
 \bottomrule
 \end{tabular}
 \caption*{References: $^1$~\citet{Woods2005}. $^2$~\citet{Weisskopf2002,ChandraX-rayCenter2013}. $^3$~\citet{Ehle2003}. $^4$~\citet{DenHerder2001}. $^5$~\citet{Bowyer1991}. $^6$~\citet{Sahnow2000,Moos2000}. $^7$~\citet{Kondo1989}.}
 \label{tab:instr}
 \end{table}

Elsewhere in the literature, $\epsilon$ Eridani has commonly been used as an analogue of the Hot Jupiter host star HD189733, the two stars being of similar type, metallicity and age \citep[e.g.][]{Moses2011,Venot2012}. AD Leonis has been used in previous studies of habitable planets \citep[e.g.][]{Tarter2007}. These are active stars that undergo frequent flaring. Flares in M dwarf stars are understood to be typically impulsive events, intense but of short duration \citep{Sanz-Forcada2002,Loyd2014}, most lasting on the order of minutes. However, it should be noted that long-lived (of order hours to days) flaring events have been detected in these stars -- one such event in particular was observed from AU Microscopii by EUVE \citep{Katsova1999}. Although these events are important, we shall not consider them any further here. We leave the study of effects of stellar time variability on EGP atmospheres to future work.

\begin{table*}
\centering
\caption{Stellar properties: spectral type, effective temperature $T_{\text{eff}}$, stellar radius $R$, distance $d$, line-of-sight hydrogen column density $N_{\text{H}}$ and age.}
\begin{tabular}{lcccccc}
\toprule
 & Spectral & $T_{\text{eff}}$ & $R$ & $d$ & $\text{log}\left(N_{\text{H}}\right)$ & Age \\
 & type & [K] & [$R^{\odot}$] & [pc]  & [cm$^{-2}$] & [Myr] \\
 \midrule
 $\epsilon$ Eri & K2V & 4900$^1$ & $0.74\pm 0.01^2$ & 3.2$^5$ & 17.8$^7$ & 730$^8$\\
 AU Mic & M1Ve & 3720$^1$ & $0.68\pm 0.17^3$ & 9.9$^6$ & 18.2$^7$ & 12$^8$\\
 AD Leo & M4.5Ve & 3370$^1$ & $0.41\pm 0.08^4$ & 4.9$^5$ & 18.5$^7$ & 25$^9$\\
 \bottomrule
 \end{tabular}
 \caption*{References: $^1$~\citet{Wright2003}, $^2$~\citet{Baines2012}, $^3$~Range of values from \citet{PasinettiFracassini2001,Wright2011,Houdebine2012,Messina2010,Rhee2007}, $^4$~Range of values from \citet{Reiners2009,Morin2008,PasinettiFracassini2001,Rutten1987,Wright2011}, $^5$~\citet{VanLeeuwen2007}, $^6$~\citet{Jenkins1952}, $^7$~\citet{Redfield2008}, $^8$~\citet{Rhee2007}, $^9$~\citet{Shkolnik2009}.}
 \label{tab:stellar_prop}
 \end{table*}

\subsection{Coronal model} \label{sec:xexo}
We use stellar coronal models \citep{Sanz-Forcada2011} to obtain XUV synthetic spectra for the three stars of interest. The thermal structure of each star's corona and transition region is constructed using an Emission Measure Distribution (EMD), which represents the quantity of emitting material at a given temperature in the stellar atmosphere. A line-based method is employed in determining the EMD, whereby individual emission line fluxes are measured in the X-ray, EUV and FUV using the stellar observatories listed in Table \ref{tab:instr}. X-ray and EUV line fluxes are measured from spectra reduced following the standard procedures described in \citet{Sanz-Forcada2003}, while FUSE fluxes are obtained from \citet{Redfield2002} and IUE spectra are downloaded from the MAST database. The obtained EMD is then constructed in such a way as to minimise the difference between the synthetic and observed line fluxes; this process is detailed further in \citet{Sanz-Forcada2003}. By combining the knowledge of this EMD with the abundances of each element and the APED (Astrophysical Plasma Emission Database) atomic model \citep{Smith2001}, the spectral energy distribution of stellar emissions in the XUV can be constructed.

For $\epsilon$ Eri, the EMD used in this study is a combination of that presented in \citet{SanzForcada2003}, which is constructed using EUV coronal lines (from EUVE observations) and FUV transition region lines (from IUE observations), with that from \citet{Sanz-Forcada2004}, which uses X-ray coronal lines obtained from Chandra. In the case of AD Leo, \citet{Sanz-Forcada2002} constructed an EMD using EUV coronal lines. For the present study, the EMD for AD Leo has been updated to also include lower temperature FUV transition region lines as well as higher temperature X-ray coronal lines: emission line temperatures now span $\text{log}(T_e) = 4.5 - 7.5$. The EMD for our final star, AU Mic, is constructed following the same procedure as for $\epsilon$ Eri and AD Leo, and is based upon emission line measurements in the X-ray, EUV and FUV.

\begin{figure}
	\centering
	\includegraphics[width=0.48\textwidth]{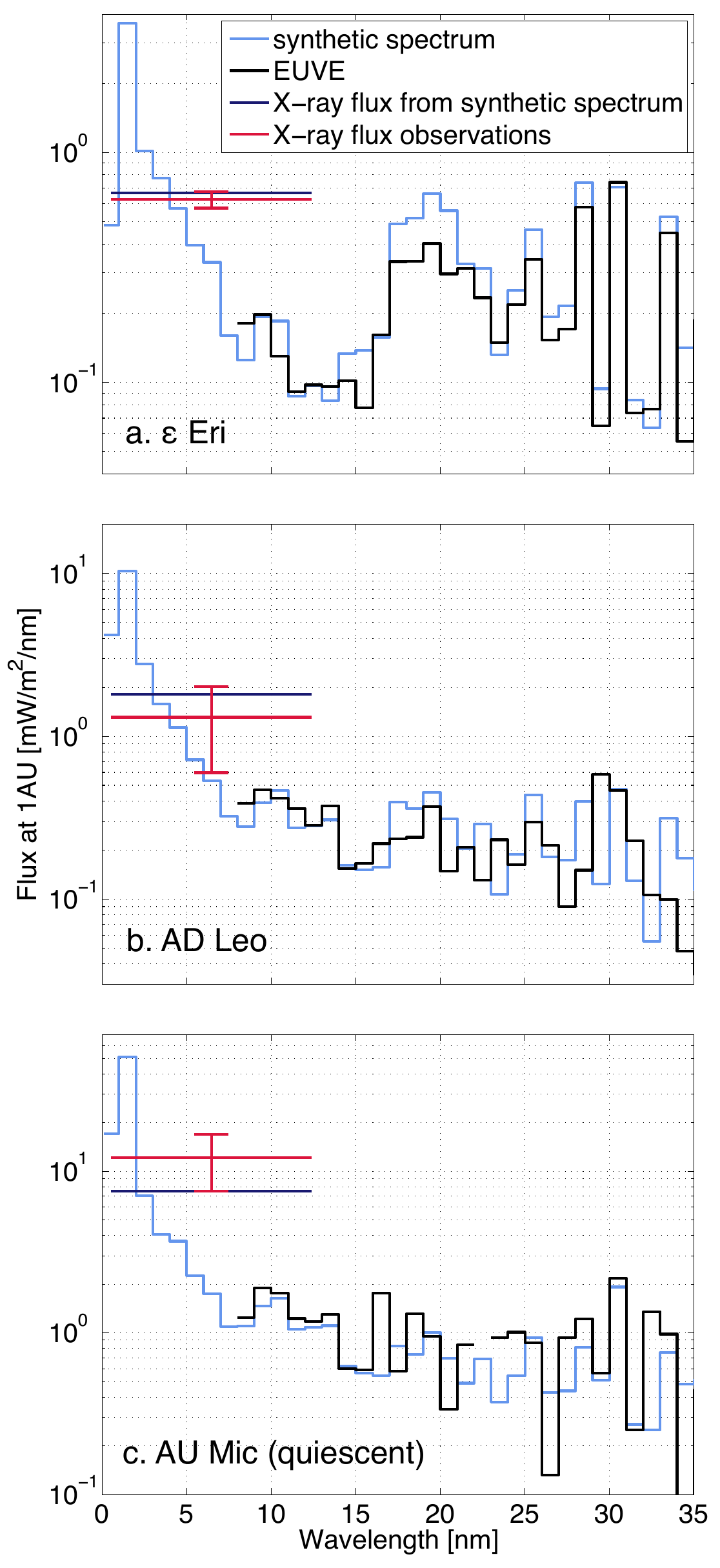}
	\caption{Comparison of synthetic spectra to X-ray and EUV observations. The light blue histograms show the synthetic spectra. The black histograms are observations from the EUVE instrument; these are co-added spectra, weighted by exposure time, for the same observations as listed in Table~\ref{tab:stellar_lum}. The red points show the range of X-ray observations in the ROSAT band from \citet{Wright2011,Schmitt2004,Lopez-Santiago2009}. To compare these ROSAT-band X-ray observations (in red) to the synthetic spectra (in light blue), the flux from each synthetic spectrum has been integrated over the same wavelength band ($\lambda \in [0.517;12.4]$~nm) as the X-ray observations; these are plotted in dark blue.}
	\label{fig:x_exo_obs}
\end{figure}

The synthetic spectra obtained have been compared to measurements in the XUV and FUV. For example, in the case of $\epsilon$ Eri, \citet{Linsky2014} compare flux levels in different wavebands between their own model, based upon Lyman $\alpha$ intensities, the \citet{Sanz-Forcada2011} coronal model -- used in this study -- and EUVE and FUSE observations. There is good agreement between both models and the observations in all the wavebands that \citet{Linsky2014} consider, i.e.\ from 10~nm to 117~nm. Note that the \citet{Linsky2014} article contains an erroneous listing in Table 6 for the X-exoplanets fluxes (those from the \citet{Sanz-Forcada2011} coronal model) between 91.2~nm and 117~nm. Indeed, this should read $\text{log}\left(f(\Delta\lambda)/f(\text{Ly}\alpha)\right) = -1.06$. This then compares very well with the observation from FUSE quoted as $\text{log}\left(f(\Delta\lambda)/f(\text{Ly}\alpha)\right) = -1.122$.

We now compare in more detail flux levels from the synthetic spectra to measurements from the ROSAT and EUVE instruments -- over the wavelength range (XUV) and at the resolution ($\Delta\lambda = 1$~nm) we use in our planetary atmosphere model. Table~\ref{tab:stellar_lum} provides the luminosity values in the X-ray and EUV wavebands for all three stars, comparing observations with results from the coronal model. The different spectral distributions observed and modelled are plotted in Fig.~\ref{fig:x_exo_obs}. The observational X-ray flux and luminosity limits provided correspond to the range of values found in the literature for ROSAT observations - the variation in X-ray emission between different observations being indicative of stellar activity. For each star, observations from the EUVE observatory have been co-added and weighted according to exposure time. ISM absorption is corrected for by applying a factor of $\text{exp}(\tau)$ to the stellar spectral irradiance, with $\tau$ being the optical depth of the column of ISM between the star and the observatory,
\begin{equation}
\tau(\lambda) = \sum_i \sigma^{\text{abs}}_i(\lambda) N_i \,,
\end{equation}
where $\sigma^{\text{abs}}_i$ is the photo-absorption cross-section of species $i$ and $N_i$ is the column density of species $i$ in the ISM, as seen from Earth. We consider that the ISM is composed of H and He, with $N_{\text{He}}=0.1N_{\text{H}}$ \citep{Spitzer1978}. Hydrogen column densities are from \citet{Redfield2008} and are given in Table~\ref{tab:stellar_prop}.

X-ray fluxes derived from the synthetic spectra over the ROSAT band (0.1 -- 2.4 keV) fall within the range of values found in the literature for all three stars, (see Fig.~\ref{fig:x_exo_obs} and Table~\ref{tab:stellar_lum}). Note that the synthetic spectrum computed for AU Mic in Fig.~\ref{fig:x_exo_obs}c is that of the star in a quiescent state -- i.e., flare events have been removed from the observations used in the construction of the EMD. Hence the integrated X-ray flux from this synthetic spectrum matches the lower boundary of X-ray observations.

The synthetic spectra also match EUVE measurements to within observational uncertainties. In terms of integrated flux between 8~nm and 35~nm, the relative difference between all co-added EUVE observations for each star and the synthetic spectra is 25\% for $\epsilon$ Eri, 5.9\% for AD Leo and 17\% for quiescent AU Mic. One cause of these discrepancies is the non-simultaneity of the different observations used in constructing the stellar EMDs. Indeed stellar X-ray flux can vary by at least a factor of 2 over the course of an activity cycle \citep{Ribas2005}. Nevertheless, the differences found between synthetic and EUVE spectra are still within the error bars for the EUVE measurements. As can be seen in Fig.~\ref{fig:x_exo_obs}, the best match in terms of spectral energy distribution between the synthetic spectra and EUVE observations is $\epsilon$ Eri. The discrepancies between 17 and 21~nm are caused by a problem in the AtomDB atomic database at these wavelengths. Despite these differences, we consider that the synthetic spectra are our `best guess' of the stellar spectra over the entire XUV range at this time.

\begin{table*}
\centering
\caption{Stellar luminosities [$10^{21}$ Watts]. The ranges given are the extremum values from the set of observations. The separation of quiescent and flaring states applies to the synthetic AU Mic spectra only. ROSAT measurements are taken from \citet{Wright2011,Schmitt2004,Lopez-Santiago2009}. The following EUVE observations are considered (same as in Fig.~\ref{fig:x_exo_obs}): for $\epsilon$ Eri: 22 Oct.\ 1993, 31 Aug.\ 1995, 5 Sept.\ 1995; for AD Leo: 5 Apr.\ 1999, 9 Apr.\ 1999, 17 Apr.\ 1999, 25 Apr.\ 1999, 6 May 1999; for AU Mic: 22 July 1993, 12 June 1996.}
\begin{tabular}{llcc|cc}
\toprule
        &       & \multicolumn{2}{c|}{ROSAT X-ray (0.517 -- 12.4~nm)} & \multicolumn{2}{|c}{EUVE (8 -- 35~nm)} \\
        &       & observations & synthetic & observations & synthetic \\
\midrule
\multicolumn{2}{l}{$\epsilon$ Eri} & $1.92$ -- $2.26$ & $2.23$ & $1.67$ -- $1.74$ & $2.16$ \\
\multicolumn{2}{l}{AD Leo}         & $2.00$ -- $6.76$ & $6.08$ & $1.11$ -- $2.63$ & $2.04$ \\
\multirow{2}{*}{AU Mic} & quiescent& \multirow{2}{*}{$25.1$ -- $56.2$} & $25.1$ & \multirow{2}{*}{$5.75$ -- $27.2$} & $6.00$ \\
 & flaring & & $252$ & & $24.7$ \\
 \bottomrule
 \end{tabular}
 \label{tab:stellar_lum}
 \end{table*}

\subsection{Scaling of the solar spectrum} \label{sec:scalings}
Due to the difficulties in measuring EUV fluxes for any stars other than the Sun (see Section \ref{sec:obs_diff}), most studies of energy deposition in exoplanetary thermospheres use solar spectra uniformly enhanced in the XUV, in place of stellar spectra. The entire XUV band is usually scaled according to the ratio of stellar to solar X-ray luminosity, $L_X^*/L_X^{\odot}$ \citep[e.g.,][]{Penz2008,Tian2009}. However, it is generally not valid to scale the EUV part of the solar spectrum using the same factor as for the X-ray band, at least for stars of different age and spectral type to the Sun. In the context of the `Sun in Time program', \citet{Ribas2005} used solar proxies of different ages, and found that power laws can be derived for the evolution of solar flux with time in different wavelength bands of the XUV. Indeed, as stars age, they lose angular momentum through frozen-in magnetic fields in the stellar wind and so progressively \mbox{spin-down}. Since coronal emissions are linked to the star's magnetic activity, these emissions diminish as the stellar dynamo declines. \citet{Ribas2005} showed that solar X-ray emissions decay faster than EUV emissions and, more generally, that higher energy solar emissions decay faster than lower energy emissions. It is likely that in other low-mass star types a similar process occurs. Indeed, \citet{Sanz-Forcada2011} determined the decay with time of the EUV and X-ray emissions for a selection of dwarf stars of various spectral types and confirmed different decay rates for X-ray and EUV emission.

\begin{figure*}
	\centering
	\includegraphics[width=0.8\textwidth]{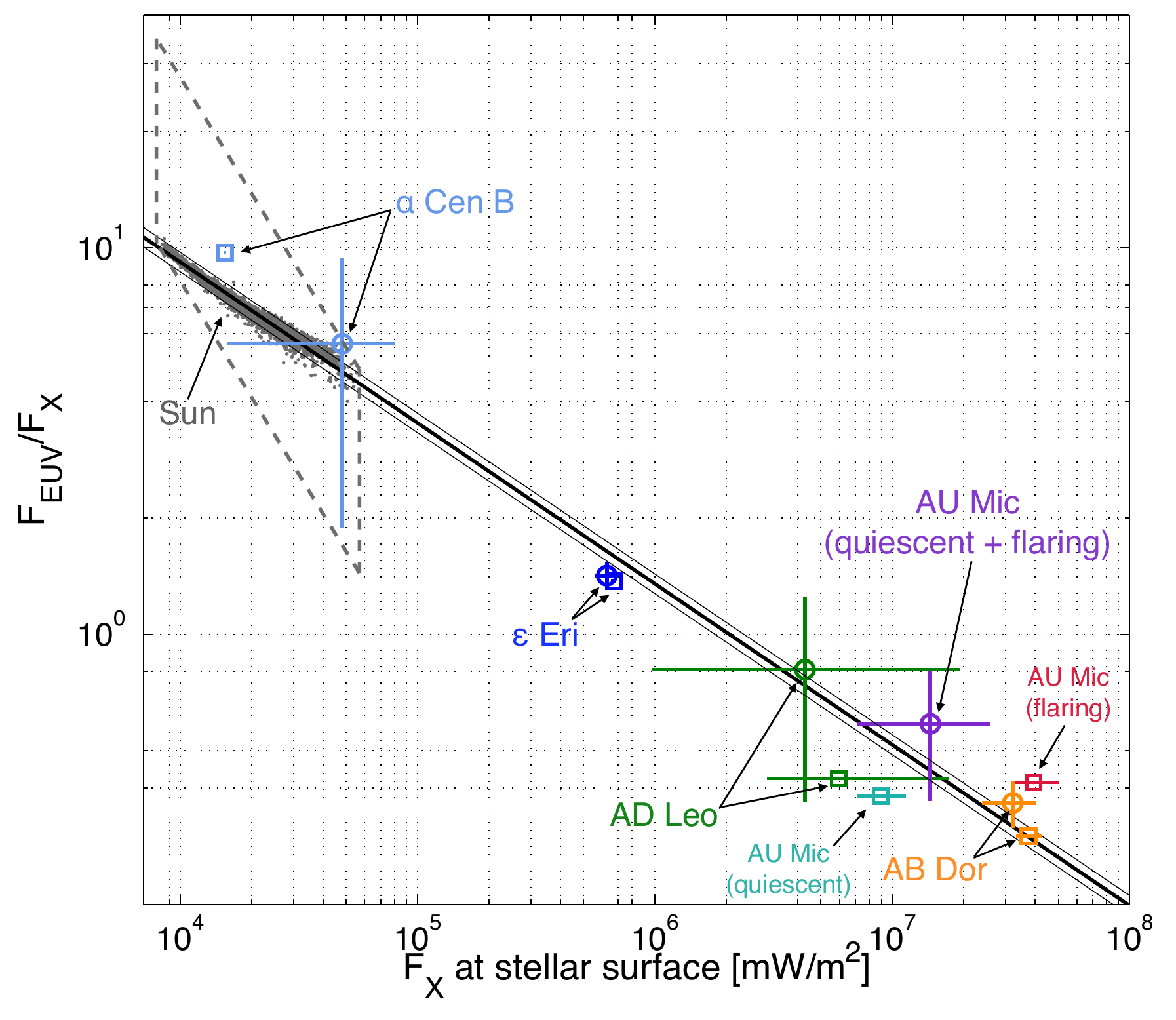}
	\caption{$F_{\text{EUV}}$-to-$F_{\text{X}}$ ratio as a function of $F_{\text{X}}$ (where $F$ is the energy flux at the star's surface) for the Sun over the course of a solar cycle (grey points) and for the stars $\alpha$ Cen B (light blue), $\epsilon$ Eri (dark blue), AD Leo (green), AU Mic (quiescent synthetic case in cyan; flaring synthetic case in red and observations in purple) and AB Dor (orange). Square markers correspond to points determined from the synthetic spectra. Open circles represent points calculated using observations for $F_{\text{X}}$ and synthetic spectra for $F_{\text{EUV}}$. The ranges indicated by horizontal / vertical bars include both variability in the X-ray observations due to stellar activity and uncertainties in the stellar radii (as given in Table \ref{tab:stellar_prop}). For AU Mic, the vertical range indicated in purple also contains variations in $F_{\text{EUV}}$, originating from the quiescent and flaring synthetic spectrum determined for this star. Solar observations are obtained from daily TIMED/SEE measurements between 30 May 2002 and 16 Nov 2013, each grey dot representing a daily averaged observation. The thick black line is a power law fitted to the solar observations (see Equation \ref{eqn:powerlaw}), 95\% confidence intervals are represented by thin black lines. The area delimited by grey dashed lines represents the largest possible extent of solar points if X-ray and EUV fluxes are taken at different times during the solar cycle.}
	\label{fig:activity_power_law}
\end{figure*}

We have derived a new power law describing the variation of stellar EUV flux as a function of X-ray flux in the ROSAT band, based on an extrapolation of emissions during the Sun's activity cycle (see Fig.~\ref{fig:activity_power_law}). To derive this scaling law, we used daily measurements between 2002 and 2013 of solar X-ray and EUV emissions, obtained from the TIMED/SEE instrument (see Table~\ref{tab:instr}), capturing a full solar cycle. Thus, we obtain the following power law (as plotted in Fig.~\ref{fig:activity_power_law}):
\begin{subequations}\label{eqn:powerlaw}
\begin{align}
\frac{F_{\text{EUV}}}{F_{\text{X}}} = 425\, \left(F_{\text{X}}\right)^{-0.42}
\label{eqn:powerlaw_pwrform}
\end{align}
\text{or, rearranging:}
\begin{align}
\text{log}F_{\text{EUV}} = 2.63 + 0.58\,\text{log}F_{\text{X}} \,,
\label{eqn:powerlaw_logform}
\end{align}
\end{subequations}
where $F$ is the stellar surface flux in mW/m$^2$. Since we are comparing stars of different spectral types, using surface fluxes rather than luminosities removes effects due to the size of the star and generally leads to better agreement over a large spectral range.

We find that more active -- and hence younger -- stars, have a lower $F_{\text{EUV}}/F_{\text{X}}$ ratio, which is consistent with the findings of \citet{Ribas2005} and \cite{Sanz-Forcada2011}. For each of the low-mass stars (other than the Sun), we use predictions from the coronal model (see Section~\ref{sec:xexo}) to obtain EUV fluxes. Thus, we have only chosen stars for which we have well constrained EMDs: in addition to $\epsilon$ Eri, AD Leo, and AU Mic, we have added $\alpha$ Cen B and AB Dor. As shown in Fig.~\ref{fig:activity_power_law}, the EUV-to-X-ray flux ratio of these stars is in good agreement with the solar behaviour, as described by the above power law (Equation \ref{eqn:powerlaw}). Note, however, that the EUV and X-ray fluxes, for all stars but the Sun, were not measured contemporaneously and might thus represent different activity levels. To illustrate the effect of stellar variability, we have indicated the range of values one can obtain when comparing non-contemporaneous solar measurements. This possible range of values is delimited by the grey dashed parallelogram in Fig.~\ref{fig:activity_power_law}, constructed using the most extreme flux cases: solar minimum X-ray with solar maximum EUV fluxes and vice versa. The extent of this area represents the largest possible uncertainty due to non-contemporaneous X-ray and EUV measurements for a star with an activity cycle of similar amplitude to that of the Sun. As such, it is most likely to be an overestimation of this effect for the other stars represented in Fig.~\ref{fig:activity_power_law}. 

To determine the response of an EGP thermosphere to irradiation by a scaled solar spectrum versus the `true' stellar spectrum (represented in this study by the synthetic spectrum from the coronal model), we determine scaling factors to apply to the solar spectrum to match emissions from each of the three stars of interest - $\epsilon$ Eri, AD Leo and AU Mic - based on both the X-ray and EUV wavebands. We define $f_{\text{X}}^*$ and $f_{\text{EUV}}^*$, the ratio of stellar-to-solar surface fluxes, for the X-ray and EUV range (respectively):
\begin{equation}
f_{\text{X}}^* = F_{\text{X}}^*/F_{\text{X}}^{\odot} \,,
\end{equation}
\begin{equation}
f_{\text{EUV}}^* = F_{\text{EUV}}^*/F_{\text{EUV}}^{\odot} \,,
\end{equation}
and $\alpha$ the ratio of a given star's EUV to X-ray flux, is given by
\begin{equation}
\alpha = F_{\text{EUV}}/F_{\text{X}} \,,
\end{equation}
where $F$ is the flux at the surface of the star (*) or the Sun ($\odot$). We determine $f_{\text{X}}^*$ using the X-ray luminosities given in Table~\ref{tab:stellar_lum} and the radius measurements from Table~\ref{tab:stellar_prop}. $\alpha^*$ can be obtained either from using a stellar coronal model (such as that described in Section~\ref{sec:xexo}) or by inserting the measured X-ray flux into Equation~\ref{eqn:powerlaw}. Once $f_{\text{X}}^*$ and $\alpha^*$ are known, $f_{\text{EUV}}^*$ can be calculated: 
\begin{equation}
f_{\text{EUV}}^* = f_{\text{X}}^*\alpha^*/\alpha^{\odot}.
\end{equation}
Here, $\alpha^{\odot}$ is derived using TIMED/SEE observations from January 2013; we take $\alpha^{\odot}=6.1$. The ratio of stellar-to-solar luminosity can be obtained by multiplying $f^*$ by a factor $(R^*/R^{\odot})^2$. Thus $L_{\text{X}}^*/L_{\text{X}}^{\odot}=f_{\text{X}}^*\left(R^*/R^{\odot}\right)^2$ is the factor by which to scale the solar luminosity to match a given star's X-ray luminosity and $L_{\text{EUV}}^*/L_{\text{EUV}}^{\odot}=f_{\text{EUV}}^*\left(R^*/R^{\odot}\right)^2$, to match the star's integrated EUV luminosity. The values of these ratios and scaling factors can be found in Table~\ref{tab:scalings} for the three stars of interest.

\begin{figure*}
	\centering
	\includegraphics[width=0.94\textwidth]{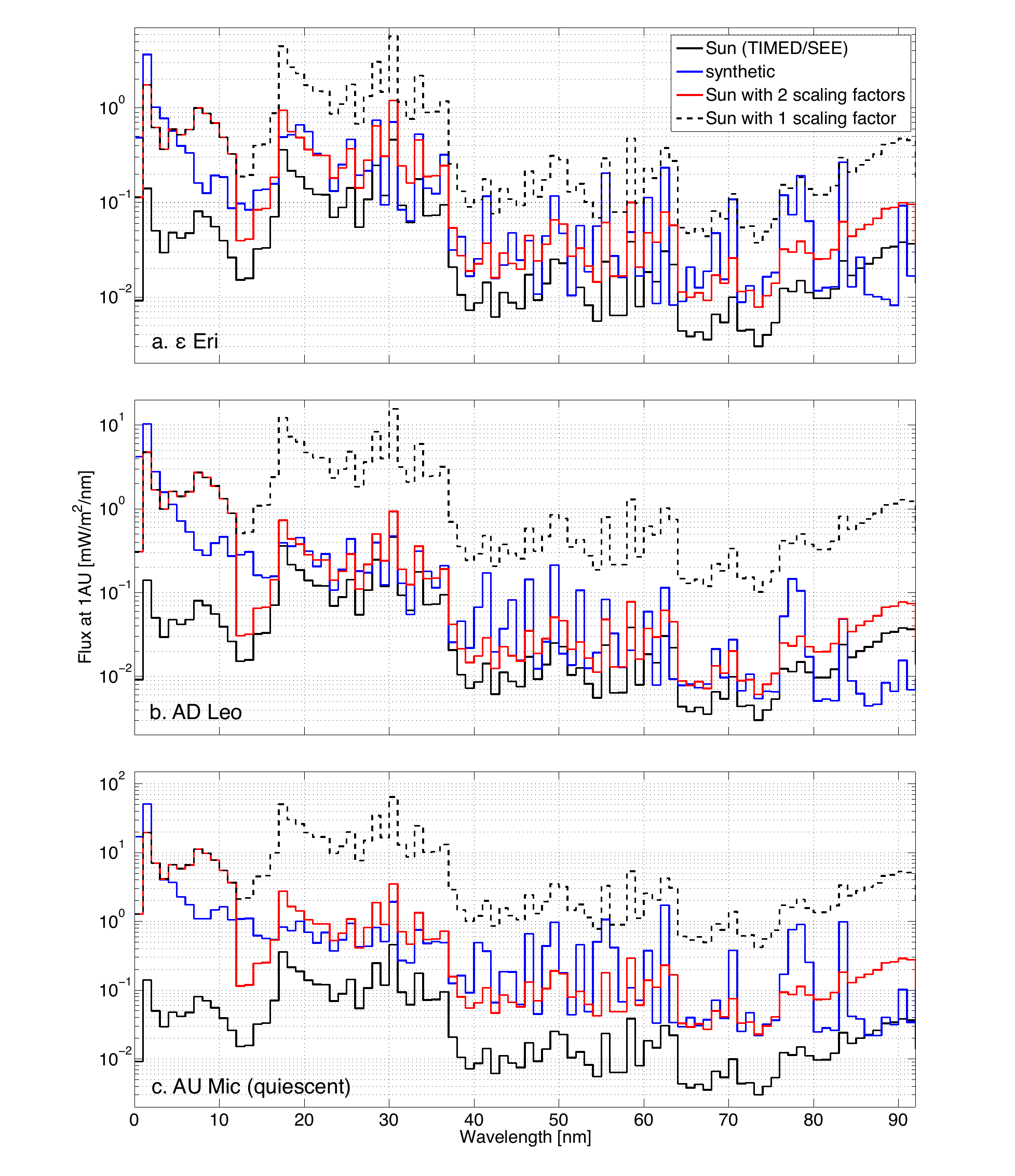}
	\caption{Comparison of synthetic spectra to scaled solar spectra. The solar spectrum used (plotted in a solid black line) is the daily averaged observation from TIMED/SEE on 14 January 2013. This solar spectrum is scaled in two different ways. Firstly, shown in dashed black lines, the solar spectrum is scaled according to a single scaling factor applied to the entire represented wavelength range and derived in such a way that the scaled solar flux in the X-ray band matches that of the synthetic spectrum. Secondly, represented in red, the solar spectrum is scaled using one scaling factor for the X-ray band and another for the EUV such that the integrated flux of the scaled spectrum matches that of the synthetic spectrum over the entire represented wavelength range (XUV). These scaled spectra, derived for each star, are to be compared to the synthetic spectra, plotted in blue.}
	\label{fig:x_exo_scaledsun}
\end{figure*}

Fig.~\ref{fig:x_exo_scaledsun} compares synthetic spectra for $\epsilon$ Eri, AD Leo and AU Mic (in blue) to scaled solar spectra using the scaling factors from Table~\ref{tab:scalings}. Two scaled solar spectra are constructed for each star. The first (dashed black line) is based on just one scaling factor: the entire XUV region is scaled using the star's X-ray luminosity alone (scaling factor of $f_{\text{X}}^*\left(R^*/R^{\odot}\right)^2$ for wavelengths between 0.1~nm and 92~nm). For the second (in red), separate scaling factors for the X-ray ($f_{\text{X}}^*\left(R^*/R^{\odot}\right)^2$ for wavelengths between 0.1~nm and 12~nm) and EUV ($f_{\text{EUV}}^*\left(R^*/R^{\odot}\right)^2$ for wavelengths between 12~nm and 92~nm) regions are used. The non-scaled solar spectrum is shown, for comparison, as a solid black line. Constructing the scaled spectra using values of $F_{\text{X}}^*$ and $F_{\text{EUV}}^*$ from the coronal model (rather than using Equation \ref{eqn:powerlaw}), allows us to assess solely the effects of different spectral energy distributions (SEDs) on the deposition of stellar radiation in upper planetary atmospheres -- the integrated flux in the XUV being conserved, by construction, between the synthetic and the scaled solar spectra (using two scaling factors).

Scaling the entire XUV region based on $f_{\text{X}}^*$ -- as has been done in previous upper planetary atmosphere studies -- gives a large overestimate of the stellar energy output in the EUV wavelength band for active stars (see black dashed line in Fig.~\ref{fig:x_exo_scaledsun}). For the case of $\epsilon$ Eri, the solar spectrum scaled using just an X-ray scaling factor gives a flux at 1~AU of 57~mW/m$^2$ integrated over 0.1 -- 92~nm, compared to 18~mW/m$^2$ predicted by the coronal model. The difference is even larger for the two other, more active stars. The X-ray scaling method gives 155~mW/m$^2$ and 642~mW/m$^2$ at 1~AU, compared to 30~mW/m$^2$ and 119~mW/m$^2$ predicted by the coronal model, for AD Leo and AU Mic (quiescent), respectively.

Quite significant differences are present between the spectral shapes of the Sun and the other stars. Most noticeably, there is a large energy excess in the scaled solar spectra between 5 and 12~nm and a deficit between 12 and 16~nm. The effect these differences in spectral shape have on exoplanetary atmospheres is assessed in Section \ref{sec:stellar_dep}.

\begin{table*}
\centering
\caption{Surface flux and luminosity ratios for the different stars. The solar fluxes are obtained from the TIMED/SEE daily average observation from 14 January 2013. X-ray fluxes for the other stars ($\epsilon$ Eri, AD Leo and AU Mic) are those used in the coronal model and the EUV fluxes are obtained from the resulting synthetic spectra.  Only the quiescent case for AU Mic is listed. The parameters given are: $f_{\text{X}}^* = F_{\text{X}}^*/F_{\text{X}}^{\odot}$; $f_{\text{EUV}}^* = F_{\text{EUV}}^*/F_{\text{EUV}}^{\odot}$; $L_{\text{X}}^*/L_{\text{X}}^{\odot}=f_{\text{X}}^*\left(R^*/R^{\odot}\right)^2$; $L_{\text{EUV}}^*/L_{\text{EUV}}^{\odot}=f_{\text{EUV}}^*\left(R^*/R^{\odot}\right)^2$; $\alpha = F_{\text{EUV}}/F_{\text{X}}$ (see text, Section \ref{sec:scalings}, for more details).}
\begin{tabular}{lcccccc}
\toprule
& $F_{\text{X}}$ (surface flux) & $f_{\text{X}}^*$ & $L_{\text{X}}^*/L_{\text{X}}^{\odot}$ & $f_{\text{EUV}}^*$ & $L_{\text{EUV}}^*/L_{\text{EUV}}^{\odot}$ & $\alpha$ \\
& [mW/m$^2$] &  &  &  &  &  \\
\midrule
Sun & $2.96\times 10^4$ & 1 & 1 & 1 & 1 & 6.13 \\
$\epsilon$ Eri & $6.70\times 10^5$ & 22.7 & 12.4 & 4.72 & 2.58 & 1.28\\
AD Leo & $5.95\times 10^6$ & 201 & 33.8 & 12.1 & 2.03 & 0.370\\
AU Mic (qsc) & $8.92\times 10^6$ & 302 & 140 & 16.5 & 7.63 & 0.334\\
\bottomrule
\end{tabular}
\label{tab:scalings}
\end{table*}

\section{Model of the upper atmosphere} \label{sec:hydro_model}

We use a one-dimensional model for the thermospheres of EGPs \citep{Koskinen2013a,Koskinen2013b,Koskinen2014} to calculate the temperature, velocity and density profiles in the upper atmospheres of planets irradiated by the different stellar spectra discussed in Section 2. In all simulations we use the planetary parameters of HD209458b (radius $R_p =$~1.32~$R_{\text{Jupiter}}$, mass $M_p =$~0.69~$M_{\text{Jupiter}}$). The model solves the vertical equations of motion from the 10$^{-6}$~bar level up to the exobase for a fluid composed of H, H$_2$, and He, as well as their associated ions H$^+$, H$_2^+$, H$_3^+$, He$^+$, and HeH$^+$. The lower boundary at 10$^{-6}$~bar is assumed to correspond either to the homopause or the level at which other molecules such as H$_2$O, CO, or CH$_4$ dissociate so that these species can be excluded from the simulations \citep{Koskinen2014}. The H$_2$/H ratio at the lower boundary is in thermal equilibrium determined by the equilibrium temperature of the planet for a given orbital distance. At the upper boundary we use either Jeans or modified Jeans boundary conditions at the exobase, depending on the value of the thermal escape parameter $X$ \citep[e.g.,][]{Hunten1973}, or outflow boundary conditions for close-in EGPs under hydrodynamic escape. Following \citet{Tian2008,Tian2008a} and \citet{Koskinen2014}, we define the hydrodynamic escape or rapid escape regime as the regime where the escape of the atmosphere leads to significant (adiabatic) cooling of the upper atmosphere.

We have performed model runs at orbital distances of 0.1, 0.2, 0.5, and 1~AU, using the different stellar spectra described in Section \ref{sec:stellar_spec}, i.e., synthetic spectra for the stars $\epsilon$ Eri, AD Leo and AU Mic; a solar spectrum from TIMED/SEE measurements on 14th January 2013, and solar spectra scaled in two different ways: using either one ($f_{\text{X}}^*$) or two ($f_{\text{X}}^*$ and $f_{\text{EUV}}^*$) scaling factors to match the integrated flux from the K and M stars over different wavelength bands (see Section 2.3 for more details on the scaling of the solar spectrum). We use a fixed heating efficiency of 93\% for photoelectrons in all of our simulations. In reality the photoelectron heating efficiency depends on the spectrum of the host star and the orbital distance, and it can also change with altitude in the atmosphere \citep[e.g.,][]{Koskinen2013a}. The purpose of this work, however, is not to exactly model the temperature and density profiles around active stars, but rather to study the relative differences in EGP atmospheres resulting from differences in the assumed spectra of their host stars.

\section{Stellar energy deposition} \label{sec:stellar_dep}

\subsection{Effect of stellar radiation on the thermosphere} \label{sec:absorption_thermo}
\begin{figure}
	\centering
	\includegraphics[width=0.5\textwidth]{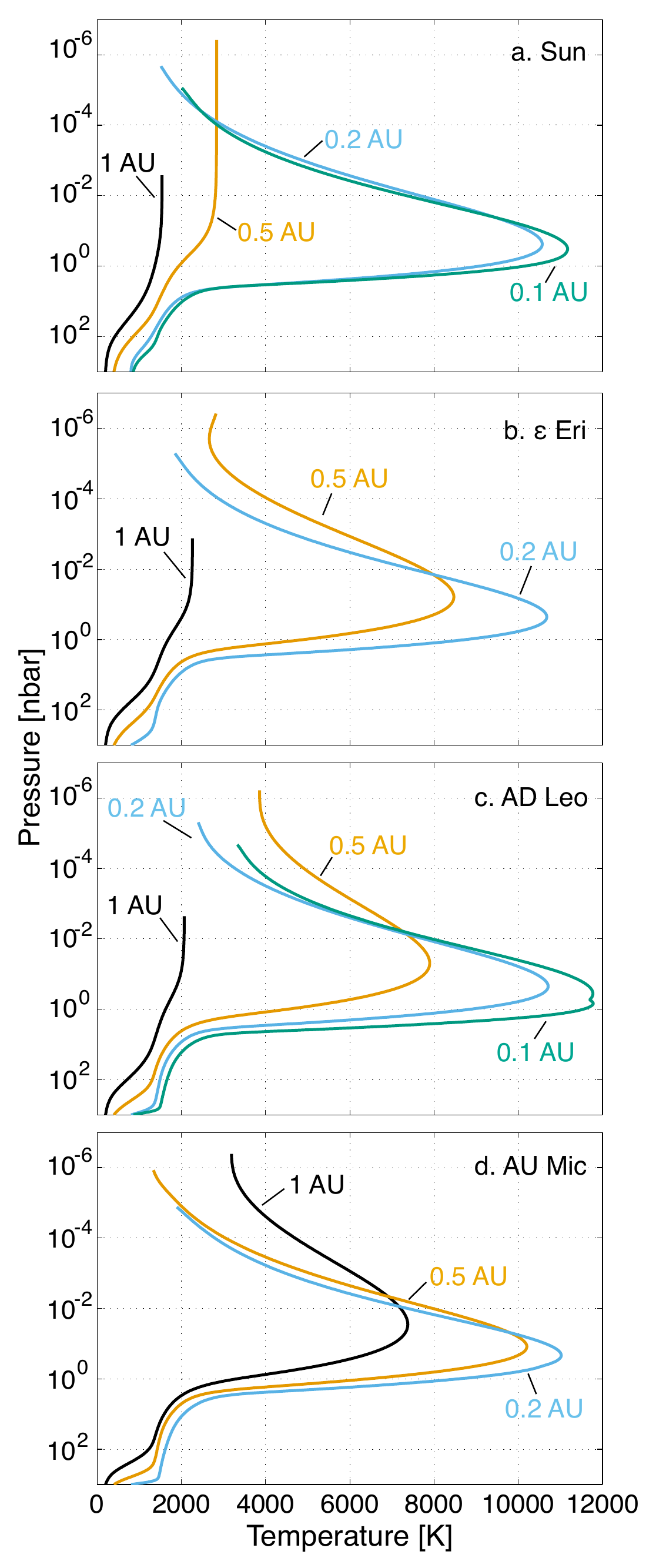}
	\caption{Temperature profiles as a function of pressure for planets orbiting the Sun (a), $\epsilon$ Eri (b), AD Leo (c) and AU Mic (d) at a distance of 1~AU (black), 0.5~AU (orange), 0.2~AU (blue) and 0.1~AU (green).}
	\label{fig:p_vs_T}
\end{figure}

Absorption of stellar XUV radiation in the thermosphere produces the temperature profiles given in Fig.~\ref{fig:p_vs_T}, for planets orbiting the Sun, $\epsilon$ Eri, AD Leo and AU Mic at various orbital distances. In the solar case, a spectrum from TIMED/SEE observed on 14 January 2013 is used; for the other stars, synthetic spectra from the coronal model described in Section \ref{sec:xexo} are used. There are two distinct regimes of EGP atmospheres depending on the stellar flux: planets orbiting far from their host star have `stable' atmospheres that undergo relatively slow Jeans escape whereas close-in planets undergo hydrodynamic escape and lose mass faster. We find that the transition between the two regimes is located between 0.2~AU and 0.5~AU for planets orbiting the Sun.

The thermospheric temperature profile for a gas giant at 1~AU, orbiting a Sun-like star (black line in Fig.~\ref{fig:p_vs_T}a) is qualitatively similar to the corresponding temperature profile in the thermosphere of the Earth. The temperature increases with altitude in the region where stellar EUV energy is deposited, principally between 100 and 0.1~nbar. Above this region, heating from stellar photons is balanced by conduction, giving an isothermal layer just below the exobase. In this case the exobase is located at 3~$\times$~10$^{-3}$~nbar. A similar picture emerges at 0.5~AU, where the atmosphere is still in the `stable' regime for giant planets orbiting the Sun (see orange line in Fig.~\ref{fig:p_vs_T}a). The enhanced stellar flux (compared to 1~AU) increases the exospheric temperature to 2800~K, up from 1500~K. The atmosphere is also significantly more extended than at larger orbital distances, the exobase now being located at a pressure of 2~$\times$~10$^{-6}$~nbar. We note that the model thermospheres at 1~AU and 0.5~AU are substantially cooled by infrared, thermal emissions from the H$_3^+$ ion around the EUV heating peak, helping to preserve the stability of the atmosphere. Such emissions have been detected repeatedly from solar system giant planets \citep[e.g.,][]{Drossart1989,Stallard2008,Miller2010,Melin2013} and recent results indicate that H$_3^+$ may in fact be the dominant ion in the low-to mid-latitude ionosphere of Saturn instead of H$^+$ \citep[e.g.,][]{Galand2009,Muller-Wodarg2012}. The effect of this cooling is visible in the temperature profile at 0.5~AU as a reduction in the temperature gradient at a pressure of around 2~nbar.

As the planet is moved closer to the host star, its atmosphere begins to undergo hydrodynamic escape -- see the 0.2~AU and 0.1~AU cases in Fig.~\ref{fig:p_vs_T}a. This is because high temperatures and increasing stellar flux lead to a high level of dissociation of H$_2$ and other molecules in the thermosphere, thus removing efficient molecular coolants, such as H$_3^+$. As a result, the temperature profile differs significantly from the Jeans escape regime: a very high peak temperature is attained -- 10,500~K at 0.2~AU; 11,200~K at 0.1~AU for planets orbiting the Sun -- followed, at higher altitudes, by a decrease in temperature due to rapid escape and the associated adiabatic cooling. We note that $\epsilon$ Eri, AD Leo and AU Mic all have higher XUV fluxes than the Sun (see Fig.~\ref{eqn:powerlaw}), so the transition from the `stable' regime to hydrodynamic escape occurs further away from the star (see Fig.~\ref{fig:p_vs_T}b-d). This transition takes place between 0.5~AU and 1~AU for planets orbiting $\epsilon$ Eri and AD Leo, and in the case of AU Mic, a gas giant orbiting at 1~AU is already in the rapid escape regime. The quantity of XUV energy emitted by the star determines the orbital distance of the transition to hydrodynamic escape.

\begin{figure*}
	\centering
	\includegraphics[width=\textwidth]{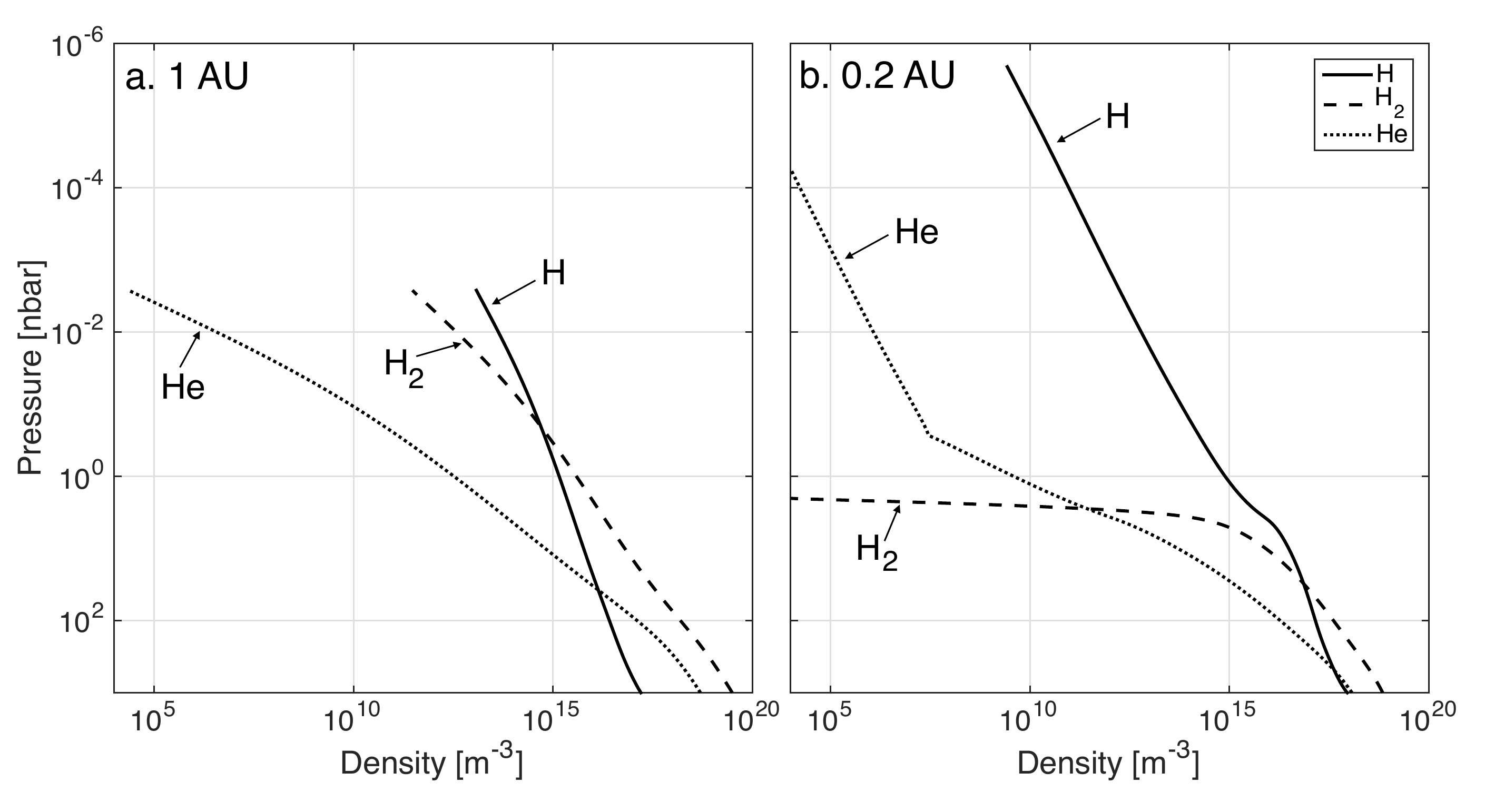}
	\caption{Densities of the neutral species with pressure calculated for a planet orbiting the Sun at 1~AU (a) and 0.2~AU (b).}
	\label{fig:p_vs_n}
\end{figure*}

Fig.~\ref{fig:p_vs_n} shows the density profiles of the three neutral species H, H$_2$, and He. Both panels represent results for a planet orbiting the Sun, at 1~AU in Fig.~\ref{fig:p_vs_n}a and at 0.2~AU in Fig.~\ref{fig:p_vs_n}b. In the `stable' atmosphere of the left panel, neutral densities drop off as a function of the molecular weight of each species. This is as expected, since we are modelling the heterosphere, where the degree of mixing is no longer sufficient to ensure constant mixing ratios with altitude and diffusive separation takes place. Thus, H$_2$ is present in significant quantities throughout the upper atmosphere, allowing for the formation of H$_3^+$ and cooling through IR emission. At the high temperatures and stellar fluxes experienced by close-orbiting planets, such as the case represented in Fig.~\ref{fig:p_vs_n}b, H$_2$ undergoes thermal and photo-dissociation (see \citet{Koskinen2010}) and is thus confined to the lower region of the model altitude grid. At higher altitudes, atomic H is the dominant species. Note that, at low pressures, the slope of the He density is the same as that of H, meaning that the two species are no longer diffusively separated. Escaping H is thus dragging He with it. Another feature to note is the sharp change in slope in the He density profile; this is due to competition between advection and diffusion timescales in the model.

\subsection{Using scaled solar spectra}
Where a full coronal model of the host star is not available, we propose using the power law provided in Equation~\ref{eqn:powerlaw} and plotted in Fig.~\ref{fig:activity_power_law} to obtain the star's EUV flux from observations in the X-ray (since EUV observations are rendered very difficult by absorption in the ISM, as discussed in Section \ref{sec:obs_diff}). In this section, we compare outputs from the thermospheric model using stellar fluxes from the coronal model and scaled solar fluxes. The solar flux is scaled in two different ways: using either 1 or 2 scaling factors, as described in Section \ref{sec:scalings}. The 1-scaling method involves scaling the entire solar XUV region by a single scaling factor to match the observed stellar X-ray flux. We shall call these spectra `X-ray scaled'. In stars more active than the Sun, this produces an overestimation of the stellar XUV flux, since the EUV flux increases at a slower rate than the X-ray flux with stellar activity (see Fig.~\ref{fig:activity_power_law}). For this reason we derived a 2-scaling method to scale the solar flux using separate scaling factors for the X-ray (using observations) and the EUV (using X-ray-to-EUV flux ratios predicted by the coronal model described in Section \ref{sec:xexo} or Equation~\ref{eqn:powerlaw}). When the EUV flux is determined using the coronal model, we shall call the resulting scaled solar spectra `EMD scaled' and when it is derived from Equation~\ref{eqn:powerlaw}, we shall call the method `parametrised scaling'. The 2-scaling case is a better approximation due to a slightly better representation of the stellar SED, but mostly because the stellar flux is conserved over the entire XUV region, which is the waveband absorbed in the planet's thermosphere.

\begin{figure}
	\centering
	\includegraphics[width=0.5\textwidth]{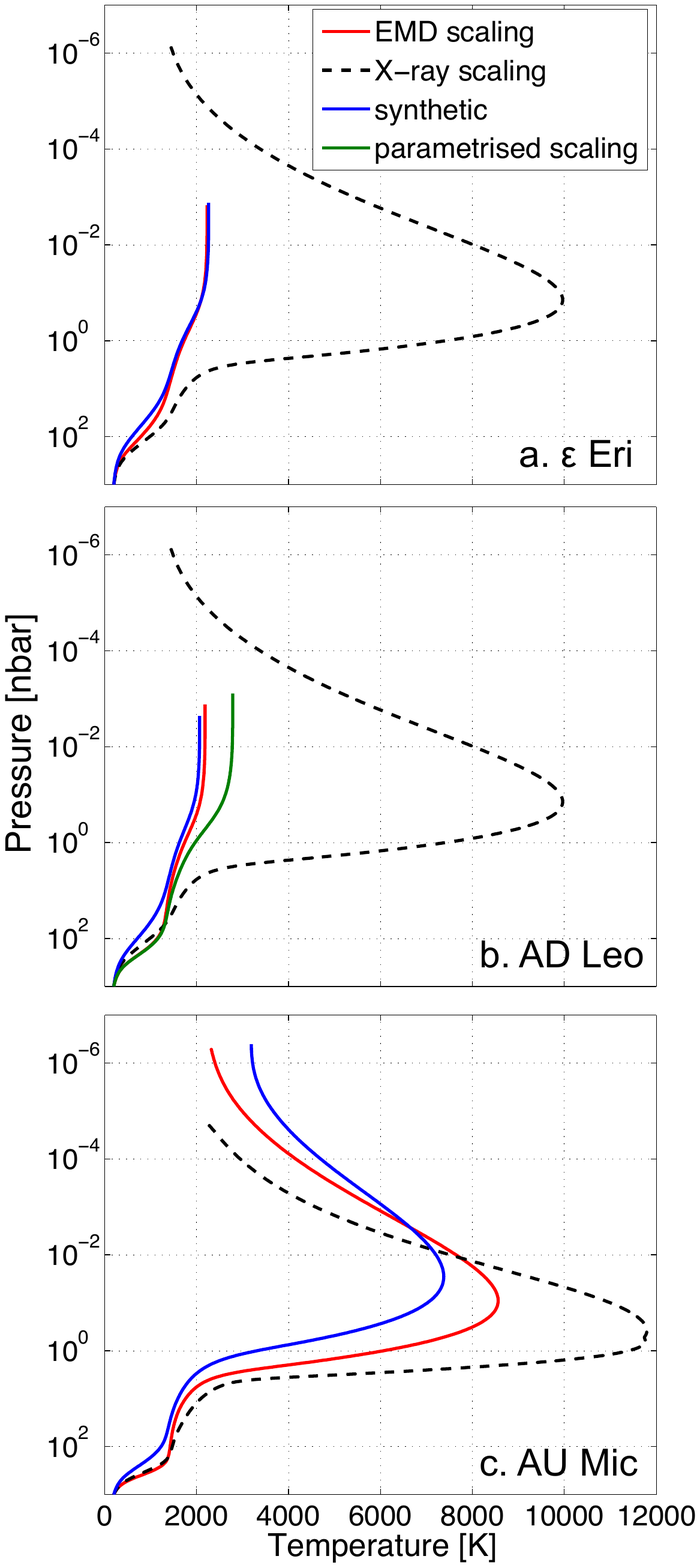}
	\caption{Temperature profiles as a function of pressure, using as energy input the scaled solar spectra given in Fig.~\ref{fig:x_exo_scaledsun} (in red, green and black dashed lines) and comparing to the synthetic case (in blue) for a planet orbiting the stars $\epsilon$ Eri (a), AD Leo (b) and AU Mic (c) at a distance of 1 AU.}
	\label{fig:p_vs_T_scal}
\end{figure}

Temperature profiles for runs at 1~AU are given in Fig.~\ref{fig:p_vs_T_scal}. At this orbital distance, planets orbiting $\epsilon$ Eri and AD Leo are in the stable regime and those around AU Mic undergo hydrodynamic escape, as can be seen by the blue lines in Fig.~\ref{fig:p_vs_T_scal}. In the cases of $\epsilon$ Eri (Fig.~\ref{fig:p_vs_T_scal}a) and AD Leo (Fig.~\ref{fig:p_vs_T_scal}b), irradiating the planet at 1~AU with the EMD-scaled spectrum (red curve) gives a temperature profile that is very close to that of a planet irradiated at 1~AU by the synthetic spectrum (in blue) derived from the coronal model. In contrast, using the X-ray scaled spectrum (dashed line) leads to a very different temperature profile. The overestimated energy input in the EUV waveband for the X-ray scaling case leads to a peak temperature of 11,800~K and enhances the escape rate by a factor of 10$^4$. This is to be compared to an exospheric temperatures of only about 2200~K predicted using the synthetic spectra.

There is a slight difference between the EMD-scaled and the synthetic spectrum in the lower portion of the altitude domain. The temperature difference between the two cases reaches around 180~K for $\epsilon$ Eri and 430~K for AD Leo, at a pressure of 100~nbar. This discrepancy is due to the additional flux between 5~nm and 12~nm when scaling the solar spectrum (see Fig.~\ref{fig:x_exo_scaledsun}). Indeed, despite the integrated flux in the X-ray and EUV bands being conserved between the synthetic and EMD-scaled spectra, the relative intensities of the different emission lines that make up the stars' spectra differ from those of the Sun. Additionally, the stellar flux in this wavelength range has a much larger effect on the temperature profiles than at longer EUV wavelengths. Indeed, the photo-absorption cross-section decreases rapidly with decreasing wavelength in the 5~nm to 12~nm range and therefore photons in this spectral range deposit their energy over a broad altitude range. This differs from longer EUV wavelengths, such as for instance, between 40~nm and 80~nm, where the photo-absorption cross-section remains relatively constant.

For a planet orbiting AU Mic (Fig.~\ref{fig:p_vs_T_scal}c), we predict that the upper atmosphere escapes hydrodynamically at all orbital distances tested, i.e., below and including 1~AU (see section \ref{sec:absorption_thermo}), as represented by the blue temperature profile determined using the synthetic spectrum for this star. In this case, there are large differences in peak temperatures between the different approaches; the synthetic spectrum gives a peak of 7380~K, the EMD-scaled spectrum gives a peak of 8560~K and the X-ray scaled spectrum gives a significantly higher peak temperature of 11,790~K. Despite this, we still obtain a far better approximation of the neutral atmosphere by using the EMD-scaled spectrum than the simple X-ray scaling -- in terms of both temperature profile and mass loss rate (see Section~\ref{sec:escape}).

Finally, we have tested the use of the parametrised scaling for the case of AD Leo (see green curve in Fig.~\ref{fig:p_vs_T_scal}b). Amongst the three stars that we include in this study, AD Leo is the one with the largest difference in F$^*_{\text{EUV}}$ between the EMD scaling -- where the flux values are taken at the green square in Fig~\ref{fig:activity_power_law} -- and the parametrised scaling -- where F$^*_{\text{X}}$ is the same as the EMD scaling, and F$^*_{\text{EUV}}$ is determined using Equation~\ref{eqn:powerlaw}, represented by the black line in Fig.~\ref{fig:activity_power_law}. Since F$^*_{\text{X}}$ is identical between the EMD and parametrised scalings, the temperature profiles at high pressure are very similar. The difference in F$^*_{\text{EUV}}$ yields a difference of around 600~K in the exospheric temperatures between the two cases with the atmospheric escape regime remaining the same. The change is therefore small compared to the X-ray scaled case (dashed line) associated with a 11,800~K peak temperature and a change in escape regime. This not only validates the parametrised approach when assessing thermospheric conditions, but also illustrates the relevance of using the parametrisation proposed in Equation~\ref{eqn:powerlaw} when the stellar EUV flux is not known.

\subsection{Atmospheric escape} \label{sec:escape}

\begin{figure}
	\centering
	\includegraphics[width=0.6\textwidth]{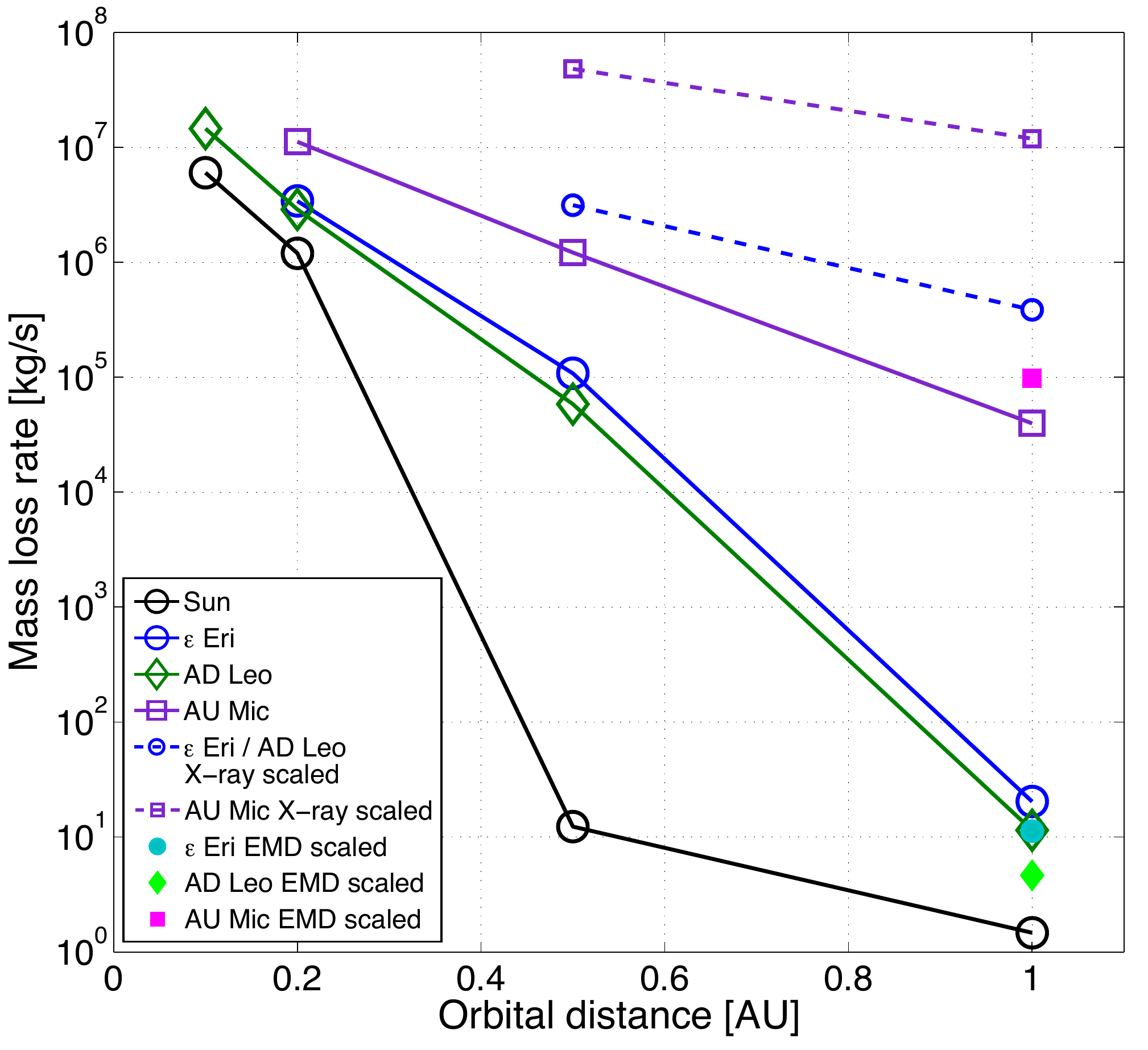}
	\caption{Mass loss rates as a function of orbital distance, for a planet orbiting different stars, as well as scaled solar cases. A TIMED/SEE spectrum is used for the Sun (in black) and synthetic spectra for the other stars (solid lines). The scaled solar cases are represented using dashed lines for the X-ray scaled spectrum and filled symbols for the EMD-scaling. The lines between points are present only to guide the eye.}
	\label{fig:X_vs_a}
\end{figure}

\begin{table*}
\centering
\caption{Mass loss rates $\dot{m}$ [kg/s] from the top of the planet's atmosphere, for planet's orbiting different host stars, at various orbital distances $a$.}
\begin{tabular}{lcccc}
\toprule
\backslashbox{Host star}{$a$} & 0.1~AU & 0.2~AU & 0.5~AU & 1~AU \\
\midrule
\textbf{Sun} & $6.0\times 10^6$ & $1.2\times 10^6$ & $12$ & $1.5$ \\
\midrule
\textbf{$\epsilon$ Eri} & & & & \\
\it{\,\,\,Synthetic} & -- & $3.4\times 10^6$ & $1.1\times 10^5$ & 20 \\
\it{\,\,\,EMD-scaled} & -- & -- & -- & 11.3 \\
\it{\,\,\,$15\times$ Sun} & -- & -- & $3.1\times 10^6$ & $3.9\times 10^5$ \\
\midrule
\textbf{AD Leo} & & & & \\
\it{\,\,\,Synthetic} & $1.5\times 10^7$ & $2.9\times 10^6$ & $5.8\times 10^4$ & 11.4 \\ 
\it{\,\,\,EMD-scaled} & -- & -- & -- & 4.6 \\
\it{\,\,\,$15\times$ Sun} & -- & -- & $3.1\times 10^6$ & $3.9\times 10^5$ \\
\midrule
\textbf{AU Mic} & & & & \\
\it{\,\,\,Synthetic} & -- & $1.2\times 10^7$ & $1.2\times 10^6$ & $4.0\times 10^4$ \\
\it{\,\,\,EMD-scaled} & -- & -- & -- & $9.8\times 10^4$ \\
\it{\,\,\,$200\times$ Sun} & -- & -- & $4.8\times 10^7$ & $1.2\times 10^7$ \\
\bottomrule
\end{tabular}
\label{tab:massloss}
\end{table*}

One of the most interesting parameters to quantify for EGPs is the escape rate, giving an idea of the lifetime of a planet's atmosphere at a given orbital distance, around a given star. Mass loss rates $\dot{m}$ for each case are given in Table~\ref{tab:massloss}. They are also shown in Fig.~\ref{fig:X_vs_a} where the two regimes of escape are visible. Thermal escape in the Jeans regime incurs mass loss rates of order 1 to 20~kg/s, whereas in the hydrodynamic escape regime, the rate jumps to $10^4$ to $10^7$~kg/s at the orbital distances that we have considered. Note that even for the largest escape rate that we have calculated -- for a planet orbiting AD Leo at 0.1~AU -- the planet's atmosphere will not be significantly depleted by this mass loss; at a rate of $1.5\times 10^7$~kg/s, the planet will only lose $4\times 10^{-4}$ of its mass in 1~Gyr. We did not perform calculations at 0.1~AU for a planet orbiting the more active star AU Mic. However, while the mass loss would be slightly higher than in an atmosphere orbiting AD Leo at the same distance, it would still be of comparable magnitude.

As can be seen in the temperature profiles described in Section \ref{sec:absorption_thermo}, the transition from a stable to a hydrodynamic escape regime occurs between 0.2 and 0.5~AU for gas-giants orbiting the Sun; between 0.5 and 1~AU for those orbiting $\epsilon$ Eri and AD Leo; and at a distance greater than 1~AU for planets orbiting AU Mic. Note that planets orbiting the K star $\epsilon$ Eri and the M star AD Leo possess very similar upper atmospheres, despite these stars having very different bolometric luminosities. When using scaled solar spectra to approximate a star's energy output, it is important to use EUV-specific scalings to obtain a good estimate of atmospheric escape. This is especially true when the atmosphere is near the transition between escape regimes. Indeed, using the EMD scaling method gives values of $\dot{m}$ that are much closer to the values based on the full synthetic spectrum (filled symbols in Fig.~\ref{fig:X_vs_a}) than the results based on the X-ray scaling (dashed lines). For instance, for a planet orbiting $\epsilon$ Eri at 1~AU, we estimate a mass loss of 20~kg/s (using the synthetic spectrum). The EMD-scaled spectrum gives a good approximation of this rate, at 11~kg/s, whereas irradiating the atmosphere with the X-ray scaled spectrum overestimates the escape rate by 4 orders of magnitude, giving $\dot{m}=3.9\times 10^5$~kg/s and an atmosphere in a different escape regime. Even in the case of a planet orbiting AU Mic at 1~AU, where the three different spectra used give an atmosphere in the fast escape regime, the EMD-scaled case gives a mass loss rate of $9.8\times 10^4$~kg/s which is far closer to the `best estimate' synthetic case rate of $4.0\times 10^4$~kg/s than the X-ray scaled spectrum (giving a mass loss of $1.2\times 10^7$~kg/s).

\section{Discussion and conclusion}
Our aim in this study has been to further the understanding of the effects of high-energy stellar radiation from low-mass stars on the upper atmospheres of extrasolar giant planets. Stellar XUV photons deposit their energy in planetary thermospheres and thus drive escape from these atmospheres. Expanding on the work of \citet{Koskinen2014} for the Sun, we confirm the existence of two distinct escape regimes in EGPs orbiting low-mass stars: a stable atmospheric regime in planets orbiting at large orbital distances and a hydrodynamic escape regime for planets orbiting close-in to their host stars. At large orbital distances, beyond the critical orbit (the transition between the two regimes), stable upper atmospheres are cooled significantly by molecular IR emissions escaping to space. In the pure H$_2$/H/He atmosphere of this study, the dominant molecular coolant is H$_3^+$. This mechanism almost vanishes at small orbital distances, where, due to the increased stellar radiation received by the planet, molecular dissociation due to thermal and photo processes increases, preventing the balancing of stellar heating by IR cooling processes.

In systems where the host star is more active than the Sun -- i.e., emits higher levels of XUV radiation -- the critical orbit is pushed further away from the star. Thus, to find stable EGPs around young stars, one has to look to larger orbital distances; in the case of a gas-giant orbiting the most active stars, such as AU Mic, the critical orbit is even beyond 1~AU. Conversely, given our results, one may be able to detect EGPs with highly expanded atmospheres at larger orbital distances from young stars than have been observed and studied up until now.

When studying upper planetary atmospheres, for example to determine escape, it is important to correctly estimate the entire XUV energy input from the host star. Indeed, the entire X-ray and EUV wavebands heat the upper atmosphere and thus drive atmospheric escape. Since the stellar flux scales differently to the solar flux in the EUV compared to the X-ray (see Fig.~\ref{fig:activity_power_law}), in order to estimate the stellar flux it is not sufficient to scale the solar XUV spectrum using one scaling factor based on the star's X-ray emissions. At least in terms of the neutral atmosphere, applying different scaling factors to the X-ray and EUV portions of the solar spectrum based on the star's integrated emissions in these wavebands is necessary and gives good results in terms of neutral temperature and density profiles. If the EUV spectrum of the host star in question is not available, we recommend using Equation~\ref{eqn:powerlaw} to estimate it based on X-ray flux observations of the star. While a two-scaling approach applied to the solar spectrum seems to be sufficient to assess thermospheric conditions, we anticipate that the stellar EUV spectrum will need to be treated more carefully when properly determining the ionised part of the upper atmosphere. This will be the subject of a follow-up paper.

There remain a lot of unknowns in the field of exoplanetary atmospheres, not least because of a lack of observations. This is both true in regards to the planetary atmospheres themselves -- there is currently very little constraint on atmospheric dynamics for example -- and in terms of the behaviour of activity cycles of low-mass stars. In recent years, however, Kepler observations have sparked a renewal of interest in stellar activity, at least in the visible, with many more results remaining to be dug out of the existing data. Stellar UV observations are currently performed with HST, but once it is decommissioned, UV capability will be lacking and there is an urgent need for a replacement mission. Proposals such as UVMag \citep{Neiner2014}, which is being submitted to ESA for consideration, are of significant value. As for planetary atmospheres, future space missions, such as NASA's JWST will give us further insight through IR transit observations. Ground-based observations have also been of great use in characterising transiting exoplanetary atmospheres \citep[e.g.,][]{Swain2010,Sing2012} and will continue to be so in the future.

\section*{Acknowledgements}
J.M.C.\ is grateful for support from the Science \& Technology Facilities Council (SFTC) through a postgraduate studentship. M.G.\ and Y.C.U.\ are partially funded by STFC through the Consolidated Grant to Imperial College London. T.T.K. acknowledges support from the National Science Foundation (NSF) grant AST 1211514. J.S.F. acknowledges support from the Spanish MINECO through grant AYA2011-30147-C03-03.

\bibliographystyle{elsarticle-harv}
\bibliography{arXiv_submission}

\end{document}